\documentstyle[preprint, aps]{revtex}

\newcommand{\be}{\begin{equation}}
\newcommand{\ee}{\end{equation}}

\newcommand{\dlt}{\delta}
\newcommand{\om}{\omega}
\newcommand{\ep}{\varepsilon}

\newcommand{\br}{{\bf r}}

\newcommand{\al}{\alpha}

\newcommand{\ra}{\rightarrow}

\newcommand{\lbd}{\lambda}
\newcommand{\prt}{\partial}

\newcommand{\vp}{\varphi}

\tightenlines

\begin{document}

\draft 

\title{Optimal trap shape for Bose gas with attractive interactions} 
\author{V.I. Yukalov$^{1,2}$ and E.P. Yukalova$^{1,3}$} 

\address{$^1$Institut f\"ur Theoretische Physik, \\
Freie Universit\"at Berlin, Arnimallee 14, D-14195 Berlin, Germany}

\address{$^2$Bogolubov Laboratory of Theoretical Physics, \\
Joint Institute for Nuclear Research, Dubna 141980, Russia}

\address{$^3$Department of Computational Physics, Laboratory of Information
Technologies, \\
Joint Institute for Nuclear Research, Dubna 141980, Russia}

\maketitle

\begin{abstract}

Dilute Bose gas with attractive interactions is considered at zero 
temperature, when practically all atoms are in Bose-Einstein condensate. 
The problem is addressed aiming at answering the question: What is the 
optimal trap shape allowing for the condensation of the maximal number 
of atoms with negative scattering lengths? Simple and accurate analytical 
formulas are derived allowing for an easy analysis of the optimal trap 
shapes. These analytical formulas are the main result of the paper.

\end{abstract}

\vskip 2cm


\section{Introduction}

Physics of dilute Bose gases is a topic of high experimental and 
theoretical interest [1--5]. One of main parameters, characterizing
the properties of these systems, is the intensity of atomic interactions 
defined by the value of the scattering length. The latter, by means of 
the Feshbach resonance techniques, can be varied in a wide range, 
including the change from positive to negative scattering lengths (see 
review articles [6,7] and references therein). Inverting the sign of the
scattering length from positive to negative means the change of effective 
atomic interactions from repulsive to attractive.

The properties of Bose systems with repulsive and attractive interactions 
are principally different. Thus, the homogeneous Bose gas with attractive 
interactions is known to be unstable for any interaction intensity [8--10].
This happens because, under attractive interactions, the sound velocity of 
the homogeneous Bose gas becomes imaginary and, respectively, the isothermal 
compressibility negative and the Bogolubov spectrum complex [8--10].

The situation is different for trapped atoms, for which attractive atomic 
interactions can be stabilized by the zero-point kinetic energy due to the 
trapping potential, provided the number of atoms does not exceed a critical 
value $N_c$. Several experiments demonstrated the existence of the 
Bose-Einstein condensate of trapped atoms with negative scattering lengths.
Thus, Bose-condensed $^7$Li with a negative scattering length was trapped 
and studied in experiments [11--13]. Bose condensates of $^{85}$Rb, with 
negative scattering lengths, were obtained by varying the latter from 
positive to negative by means of the Feshbach resonance techniques [14,15].

We may note that the Feshbach resonance techniques were also used for 
studying the instability in the dynamics of Bose-condensed gases, which 
developed owing to an abrupt change between the {\it repulsive} weak-coupling 
to strong-coupling regimes. For instance, in the experiment [16], a sample 
of $^{85}$Rb atoms was evaporatively cooled and condensed into a state of 
atoms with a {\it positive} but a very small scattering length of about 80 
$a_B$, where $a_B$ is the Bohr radius. Then, by means of a fast variation 
of the magnetic field close to the Feshbach resonance, the scattering 
length was rapidly increased to 1900 $a_B$. The dynamic instability, 
observed in this rapid increase of the {\it positive} scattering length 
has a very different physical origin [16] and will not be discussed here. 
In the present paper, we shall consider only equilibrium Bose-condensed 
gases with attractive interactions, hence, with {\it negative} scattering 
lengths.

Defining the critical number of Bose-condensed atoms, N$_c$, which could 
be loaded into a trap, requires to accomplish heavy numerical calculations, 
since there are no small parameters that could facilitate the calculational 
procedure. The critical number N$_c$ was calculated by different numerical 
methods for spherically symmetric traps [17--24] and for anysotropic traps 
[25--27]. The general understanding for achieving the better collapse-avoiding
properties is as follows. Attractive atomic interactions are stabilized by 
the zero-point kinetic energy due to the trapping potential. In order to 
minimize the effect of attractive interactions for a given number of atoms, 
it is useful to reduce the condensate density, i.e., to use traps with 
lower confinement. The zero-point energy, corresponding to the weakest 
confining trap direction, is the one relevant for determining the stability 
limits. When the radial trap frequency was held fixed, a cigar-shaped trap 
had a higher critical number mostly because doing so lowered the condensate 
density from what it would have been if the trap was spherical. Eventually,
the lowered confinement in the weaker direction resulted in too little 
zero-point energy to stabilize the atoms. Thus, there should be an optimal 
critical number as one of the trapping frequencies would be relaxed. If 
the optimal trap shape were considered under a fixed condensate density, 
then the spherical shape would be optimum. 

Although the overall physical picture, resulting from numerical calculations 
[17--27], is understandable, there are no simple analytical relations 
allowing for a not too complicated investigation of the attractive 
Bose-Einstein condensate stability under the given parameters of an 
anisotropic trap. It is the aim of the present paper to advance a novel 
theoretical approach for the problem, using which we derive approximate 
{\it analytical formulas} connecting the critical number of atoms $N_c$ 
with the trap frequencies. Though the derivation requires the usage of 
some elaborated mathematics, the final expressions are quite simple. Being 
simple, our formulas are at the same time rather accurate. Comparing them 
with the known numerical calculations [17--27], we find a good agreement, 
with a deviation within the range of the order $10\%$. No such analytical 
formulas have been obtained earlier. The derived analytical expressions 
can serve as a convenient tool for understanding and estimating the basic 
relations between the trap shape and the maximal number of trapped atoms 
with negative scattering lengths.

Throughout the paper the system of units is employed, where the Planck 
constant, $\hbar=1$, and the Boltzmann constant, $k_B=1$, are set to 
unity.

\section{Condensate Wave Function}

We consider a dilute weakly interacting Bose gas, corresponding to the 
inequality $\rho|a_s|^3\ll 1$, in which $\rho$ is the average atomic density 
and $a_s$ is the $s$-wave scattering length. For repulsive interactions, 
$a_s>0$, while for attractive interactions, $a_s<0$. The dilute Bose gas 
in the low-temperature limit becomes almost completely condensed, forming 
Bose-Einstein condensate with the number of atoms $N_0$ being approximately 
equal to the total number of atoms $N$. The condensate wave function, as 
is known [1--5], satisfies the Gross-Pitaevskii equation
\be
\label{1}
\left [ -\; \frac{\nabla^2}{2m} + U(\br) +\Phi_0|\eta(\br)|^2\right ]
\eta(\br) =\ep\; \eta(\br) \; ,
\ee
where $m$ is atomic mass, $U(\br)$ is a trapping potential, and the 
interaction strength is
\be
\label{2}
\Phi_0 = 4\pi\; \frac{a_s}{m} \; .
\ee
Equation (1) has the form of the stationary nonlinear Schr\"odinger 
equation representing the eigenvalue problem, with the eigenfunction 
$\eta(\br)$ and eigenvalue $\ep$. The confining potential is usually 
described by the harmonic potential
\be
\label{3}
U(\br) = \frac{m}{2}\; \sum_\al \; \om_\al^2\; r_\al^2 \; ,
\ee
whose effective frequencies $\om_\al$ define the anisotropy of the trap. 
The condensate wave function is normalized to the number of condensed 
atoms,
\be
\label{4}
\int |\eta(\br)|^2\; d\br = N_0 \; .
\ee
Under the considered conditions, this number is close to the total number 
of atoms, $N_0\approx N$.

The standard traps usually have the geometry of a cylinder or are almost 
cylindrical, which we assume in what follows. Then there are two
characteristic trap frequencies, the transverse, or radial, frequency
\be
\label{5}
\om_\perp \equiv \om_x = \om_y
\ee
and the longitudinal, or axial, frequency $\om_z$. The anisotropy parameter 
\be
\label{6}
\lbd \equiv \frac{\om_z}{\om_\perp}
\ee
defines the actual trap shape. When $\lbd<1$, the trap is called cigar-shaped, 
or pencil-shaped. For $\lbd=1$, the trap is spherical. And when $\lbd>1$, one 
terms the trap as disk-shaped, or pancake-shaped. Related to the characteristic
frequencies $\om_\perp$ and $\om_z$, there are two oscillator lengths
\be
\label{7}
l_\perp \equiv \frac{1}{\sqrt{m\om_\perp}} \; , \qquad l_z \equiv
\frac{1}{\sqrt{m\om_z}} \; .
\ee
One more typical length is defined through the geometric-average frequency 
$\om_0$,
\be
\label{8}
l_0 \equiv \frac{1}{\sqrt{m\om_0}}\; , \qquad \om_0 \equiv
\left ( \om_\perp^2\; \om_z\right )^{1/3} \; .
\ee
From Eqs. (5) to (8), it is straightforward to notice the relations between 
the characteristic frequencies,
\be
\label{9}
\om_0 = \lbd^{1/3} \om_\perp = \frac{\om_z}{\lbd^{2/3}} \; ,
\ee
and between the oscillator lengths,
\be
\label{10}
l_0 = \frac{l_\perp}{\lbd^{1/6}} = \lbd^{1/3} l_z \; .
\ee

Employing the cylindrical coordinates $\br=\{ r_\perp,\vp,r_z\}$, where 
$r_\perp=\sqrt{r_x^2+r_y^2}$, it is convenient to introduce the dimensionless
variables
\be
\label{11}
r \equiv \frac{r_\perp}{l_\perp} \; , \qquad
z \equiv \frac{r_z}{l_\perp} \; .
\ee
The equality
\be
\label{12}
\eta(\br) \equiv \sqrt{\frac{N_0}{l_\perp^3}}\; \psi(r,\vp,z)
\ee
defines the dimensionless wave function $\psi(r,\vp,z)$ of the radial 
variable $r\in[0,\infty)$, angle variable $\vp\in[0,2\pi)$, and of the
longitudinal variable $z\in(-\infty,+\infty)$. The normalization condition 
(4) transforms to
\be
\label{13}
\int |\psi(r,\vp,z)|^2 \; rdr\; d\vp \; dz = 1 \; .
\ee
With the notation for the dimensionless energy
\be
\label{14}
E \equiv \frac{\ep}{\om_\perp} \; ,
\ee
the Gross-Pitaevskii equation (1) takes the form of the eigenvalue problem
\be
\label{15}
\hat H_{NLS}\; \psi = E\; \psi \; ,
\ee
in which the nonlinear Schr\"odinger Hamiltonian is
\be
\label{16}
\hat H_{NLS} \equiv - \; \frac{\nabla^2}{2m} + \frac{1}{2}\left (
r^2 +\lbd^2 z^2\right ) + g|\psi|^2 \; .
\ee
Here
$$
\nabla^2 = \frac{\prt^2}{\prt r^2} + \frac{1}{r}\; \frac{\prt}{\prt r} +
\frac{1}{r^2}\; \frac{\prt^2}{\prt\vp^2} + \frac{\prt^2}{\prt z^2}
$$
and the notation for the effective coupling parameter
\be
\label{17}
g \equiv 4\pi\; \frac{a_s}{l_\perp}\; N_0
\ee
is introduced.

Equation (15), with the Hamiltonian (16), clearly shows that the energy $E$ 
is a function of $g$ and $\lbd$. Real-valued solutions for the eigenvalue 
$E=E(g,\lbd)$ exist not for all parameters $g$ and $\lbd$. For some values 
of these parameters, the eigenvalue $E$ can become complex. The appearance 
of the imaginary part in the energy $E$, means that the state with this 
energy level becomes unstable, having a finite lifetime that can be 
estimated as
$$
\tau \equiv \frac{1}{|{\rm Im}\; E|} \; .
$$
The boundary on the plane $g-\lbd$, where real-valued solutions for $E$ 
disappear, defines the critical line $g_c(\lbd)$, which, through the 
relation
\be
\label{18}
g_c(\lbd) = 4\pi\; \frac{a_s}{l_\perp}\; N_c \; ,
\ee
gives the critical number of atoms $N_c$. For the number of atoms $N_0<N_c$, 
the eigenproblem (15) possesses well-defined solutions, with real-valued 
energies. But, if $N_0$ exceeds $N_c$, the energy becomes complex, which 
means that the system is unstable and desintegrates. This critical number 
follows from Eq. (18) yielding
\be
\label{19}
N_c = \frac{g_c(\lbd)}{4\pi} \left ( \frac{l_\perp}{a_s} \right ) = 
\frac{g_c(\lbd)}{4\pi} \; \lbd^{1/2} \left ( \frac{l_z}{a_s} \right )\; .
\ee
Or, according to Eq. (10), one has
\be
\label{20}
N_c = \frac{g_c(\lbd)}{4\pi} \; \lbd^{1/6} 
\left ( \frac{l_0}{a_s} \right )\; .
\ee

To find the critical line $g_c(\lbd)$, it is necessary to solve Eq. 
(15) for varied parameters $g$ and $\lbd$. Directly solving this nonlinear 
equation, with an external potential, requires to invoke cumbersome numerical 
calculations. It would certainly be desirable to obtain an explicit 
analytical expression for the critical line $g_c(\lbd)$, which would permit 
us to accomplish a straightforward investigation of Eqs. (19) and (20).

\section{Optimized Perturbation Theory}

The behaviour of the energy $E$, as a function of $g$ and $\lbd$, can be 
found by using the optimized perturbation theory. This theory, advanced in 
Ref. [28], is based on introducing into the calculational procedure, such 
as perturbation theory or iterative algorithm, control functions, which 
govern the convergence of the calculational scheme. Control functions can 
be incorporated either into the initial approximation or into the resulting 
series by changing the variables. Then, to guarantee the convergence of the
sequence of approximations, the control functions are defined by means of 
an optimization procedure. This results in the sequence of optimized 
approximants. The specific features of the optimized perturbation theory
is its high accuracy and the possibility of its usage for the problems with 
no small parameters. This theory is nowadays widely employed for various 
problems (see the review-type papers [29,30] and references therein). In
particular, it has been successfully applied for calculating the 
interaction-induced shift of the condensation temperature of dilute Bose
gas [31--36], yielding very accurate results, well agreeing with numerical 
Monte Carlo simulations [37--41].

To solve Eq. (15), we may start with the zero-order Hamiltonian
\be
\label{21}
\hat H_0 = -\; \frac{\nabla^2}{2m} + \frac{1}{2}\left ( u^2 r^2 +
\lbd^2 v^2 z^2\right )\; ,
\ee
in which $u$ and $v$ are control functions. The corresponding zero-order
energy is
\be
\label{22}
E_0 = (2n +|m|+1)\; u + \left ( k +\frac{1}{2}\right )\; \lbd \; v\; ,
\ee
in which $n=0,1,2,\ldots$ is the radial quantum number, $m=0,\pm 1,\pm 2, 
\ldots$ is the azimuthal quantum number, and $k=0,1,2,\ldots$ is the axial 
quantum number. Employing the Rayleigh-Schr\"odinger perturbation theory, 
we get a sequence of the approximants
\be
\label{23}
E_j = E_j(g,\lbd,u,v) \; .
\ee
For brevity, we do not include explicitly the dependence on the quantum 
numbers. The control functions are defined in each order through optimization 
conditions, e.g., through the variation
\be
\label{24}
\dlt E_j = \frac{\prt E_j}{\prt u}\; \dlt u +
\frac{\prt E_j}{\prt v}\; \dlt v = 0 \; ,
\ee
which gives the functions
\be
\label{25}
u_j= u_j(g,\lbd) \; , \qquad v_j= v_j(g,\lbd) \; .
\ee
Substituting these back into Eq. (23), we obtain the optimized 
approximants
\be
\label{26}
\tilde E_j \equiv E_j(g,\lbd,u_j,v_j) \; ,
\ee
with the control functions (25).

To simplify the following expressions, it is useful to introduce an 
effective coupling
\be
\label{27}
x \equiv 2g\sqrt{\lbd}\; I \; ,
\ee
where the integral
\be
\label{28}
I \equiv \frac{(|\psi|^2,|\psi|^2)}{u\sqrt{\lbd v}}
\ee
is defined through the matrix element of the wave functions $\psi$ 
corresponding to the Hamiltonian (21). Explicitly, the wave functions are
$$
\psi(r,\vp,z) =\left [ \frac{2n!\; u^{|m|+1}}{(n+|m|)!}\right ]^{1/2} \;
r^{|m|}\exp\left ( -\; \frac{u}{2}\; r^2\right ) \times
$$
$$
\times L_n^{|m|}\left (ur^2\right )\; \frac{e^{im\vp}}{\sqrt{2\pi}}\; 
\frac{1}{\sqrt{2^k k!}}\left (\frac{\lbd v}{\pi}\right )^{1/4} \;
\exp\left ( -\; \frac{\lbd}{2}\; vz^2\right )\; H_k(\sqrt{\lbd v}\; z)\; ,
$$
where $L_n^m(\cdot)$ is an associated Laguerre polynomial and $H_k(\cdot)$ 
is a Hermite polynomial. Then the integral (28) takes the form
$$
I = \frac{2}{\pi^2} \; \left [ \frac{n!}{(n+|m|)!\; 2^k\; k!}\right ]^2\;
\int_0^\infty \; x^{2|m|}\; e^{-2x}\; \left [ L_n^{|m|}(x)\right ]^4\;
dx \; \int_0^\infty \; e^{-2t^2}\; H_k^4(t)\; dt \; ,
$$
which is a number not depending on the control functions $u$ and $v$.
In the first order, we have the energy
\be
\label{29}
E_1 = \frac{p}{2}\left ( u +\frac{1}{u}\right ) + \frac{q}{4}\left (
v + \frac{1}{v}\right ) + \frac{1}{2}\; u\sqrt{v}\; x \; ,
\ee
in which the notation
\be
\label{30}
p \equiv 2n +|m|+1 \; , \qquad q \equiv (2k+1)\lbd
\ee
is used. The variation of Eq. (29) defines the control functions by the 
optimization equations
\be
\label{31}
1-\; \frac{1}{u^2} + \frac{\sqrt{v}}{p}\; x = 0 \; , \qquad
1-\; \frac{1}{v^2} + \frac{ux}{q\sqrt{v}} = 0 \; .
\ee
These equations are to be supplemented by the boundary conditions
\be
\label{32}
\lim_{x\ra 0} u = \lim_{x\ra 0} v  = 1 \; ,
\ee
requiring that, when atomic interactions are switched off, the solutions 
be recovered corresponding to a harmonic oscillator. By their meaning, 
the control functions play the role of effective frequencies, because of 
which they have to be positive. From Eqs. (31) it follows that the range 
of definition for the solutions $u$ and $v$ depends on the sign of 
atomic interactions. Thus, for repulsive interactions, when $g>0$ and 
$x>0$, these solutions are to be in the interval
\be
\label{33}
0< u < 1 \; , \quad 0 < v < 1 \qquad (x>0).
\ee
But for attractive interactions, with $g<0$ and $x<0$, one has
\be
\label{34}
u > 1 \; , \quad v> 1 \qquad (x<0)\; .
\ee
Equation (29), together with the optimization conditions (31), gives the 
optimized approximant $\tilde E=\tilde E_1$, which we shall analyze in 
what follows. This optimized approximant can be written in the form
\be
\label{35}
\tilde E= \frac{p}{u} + \frac{q}{4}\left ( v +\frac{1}{v}\right ) \; ,
\ee
in which $u$ and $v$ are the solutions to Eqs. (31).

The set of Eqs. (31) and (35) defines the control functions $u$ and $v$ 
and the optimized approximant $\tilde E$ as functions of the effective 
coupling (27) and quantum numbers (30). Explicit dependence on these 
parameters can be obtained in the limits of weak and strong coupling, 
as is demonstrated in the Appendix A.

Of special interest is the behaviour of the ground-state level, which, 
in the case of attractive interactions, becomes unstable before the higher 
excited levels. For the ground state, one has $n=m=k=0$, hence $p=1$ and 
$q=\lbd$. The integral (28) is $I=(2\pi)^{-3/2}$. The optimization 
equations (31) reduce to
\be
\label{36}
1-\; \frac{1}{u^2} +\sqrt{v}\; x = 0 \; , \qquad 
1-\; \frac{1}{v^2} + \frac{u\; x}{\lbd\sqrt{v}} = 0 \; ,
\ee
where the effective coupling (27) is
\be
\label{37}
x =\frac{2g\sqrt{\lbd}}{(2\pi)^{3/2}} \; .
\ee
The optimized approximant (35) becomes
\be
\label{38}
\tilde E = \frac{1}{u} + \frac{\lbd}{4}\left ( v + \frac{1}{v}\right )\; .
\ee
Expression (38) represents the ground-state energy with a very good accuracy 
for the whole range of the interaction strength, from zero to arbitrary large 
values. Thus, for weak coupling, Eq. (38) gives an asymptotically exact 
value
\be
\label{39}
\tilde E \simeq 1 +\frac{\lbd}{2} + \frac{g\sqrt{\lbd}}{(2\pi)^{3/2}}
\qquad (g\ra \pm 0) \; .
\ee
The accuracy slightly diminishes for increasing interactions. However, 
even in the limit of infinitely strong repulsive interactions, Eq. (38)
yields the limit
\be
\label{40}
\tilde E \simeq 0.547538\; (g\lbd)^{2/5} \qquad (g\ra\infty) \; ,
\ee
which is only $2\%$ different from the Thomas-Fermi limit that is known to 
be asymptotically exact for $g\ra\infty$. For large attractive interactions,
when $g<0$, there is no simple asymptotic form for $g\ra-\infty$, since
the system becomes unstable at a finite critical value $g_c(\lbd)$. The 
consideration of attractive interactions is more complicated and, to find 
explicitly the critical line $g_c(\lbd)$, we need to involve some more 
techniques.

\section{Attractive Atomic Interactions}

Now the consideration of the previous section will be continued with the 
specification for attractive interactions, when $g<0$ and $x<0$. First 
of all, let us note that real-valued solutions for the ground-state energy 
(38), with the control functions from Eqs. (36), exist only for $x>x_c$,
such that
\be
\label{41}
|x_c| < 1 \qquad (0\leq \lbd <\infty) \; .
\ee
The proof of this inequality is as follows. Equations (36) can be 
rearranged to the form
\be
\label{42}
\frac{\lbd^2}{x^2}\left ( 1 - |x|\sqrt{v}\right ) =
\frac{v^3}{(v^2-1)^2} \; .
\ee
As far as the right-hand side of Eq. (42) is positive, it should be that
$$
1 - |x|\sqrt{v} \geq 0 \; .
$$
From here, $|x|\leq 1/\sqrt{v}$. Since, according to Eq. (34), $v>1$, if 
$x<0$, then $|x|<1$. The latter results in Eq. (41).

Because the critical value is smaller than one, $|x_c|<1$, for any 
anisotropy parameter $\lbd\geq 0$, it is reasonable to study, first, the 
situation at asymptotically small $|x|\leq|x_c|<1$. Making the replacement 
$x=-|x|$, we may analyze the behaviour at small $|x|$ for the control 
functions as well as for the energy. Excluding $v$ in the optimization 
conditions (36), we get the equation
\be
\label{43}
x^2\left ( u^2 -1 \right )^3 u^3 - \lbd\left ( u^2 -1 \right )^4 +
\lbd\; x^4\; u^8 = 0 
\ee
for the control function $u$. Vice versa, excluding $u$, we have Eq. (42) for 
the control function $v$. Energy (38) can be represented in terms of one of 
these functions, for instance,
\be
\label{44}
\tilde E = \frac{1}{u} +\frac{\lbd}{4}\left [ \frac{(u^2-1)^2}{x^2u^4}
+ \frac{x^2u^4}{(u^2-1)^2}\right ] \; .
\ee
Using Eqs. (42),(43),(38) or (44), we can find the expansions for all 
functions in powers of $|x|$. The expansions for the control functions are
\be
\label{45}
u\simeq \sum_{k=0}^j u_k\; |x|^k
\ee
and
\be
\label{46}
v\simeq \sum_{k=0}^j v_k \; |x|^k \; .
\ee
The expansion for the energy is 
\be
\label{47}
\tilde E \simeq \sum_{k=0}^j c_k \; |x|^k \; .
\ee
In Eqs. (45) to (47), $j=0,1,2,\ldots$. The first several coefficients 
for these expansions are given in the Appendix B.

Expansions (45) to (47) have sense for asymptotically small $|x|\ll 1$. The 
value $x_c(\lbd)$, should be smaller than one, in agreement with Eq. (41).
However, $|x_c|$ is not necessarily asymptotically small. This can be 
illustrated by the spherically symmetric case, when $\lbd=1$. Then $u=v$, 
and Eqs. (31) reduce to just one equation
\be
\label{48}
x u^{5/2} + u^2 -1 = 0 \qquad (\lbd=1) \; .
\ee
The latter equation possesses real solutions only for $x>x_c$, such that
\be
\label{49}
x_c = -\frac{4}{5^{5/4}} = - 0.534992 \; ,
\ee
which is related to $g_c=-4.212958$. Note that the critical value (49) 
follows exactly from Eq. (48).

In this way, we need more general expressions for the sought functions, but 
not merely their asymptotic expansions (45) to (47). This can be achieved by 
resorting to a method of an effective summation of power series. A very 
accurate method of summation, whose mathematical foundation is based on the 
self-similar approximation theory [42--46], is the method of self-similar
factor approximants [47--49], which was shown to be more general and accurate 
than the method of Pad\'e approximants, the latter being just a particular 
case on the class of rational functions.

The self-similar factor approximant of the $j$-th order for a power 
series, say for expansion (45), has the form
\be
\label{50}
u_j^* = u_0 \prod_{i=1}^{N_j} \left ( 1 + A_i\;|x|\right )^{n_i} \; ,
\ee
in which $N_j=j/2$, when $j$ is even, and $N_j=(j+1)/2$, with $A_1$ set to 
one, if $j$ is odd. The coefficients $A_i$ and the powers $n_i$ are defined 
by the accuracy-through-order procedure, that is, by expanding the approximant
(50) in power series with respect to $|x|$ and equating in that expansion the 
same-order terms with those of series (45). In the same way the approximants 
$v_j^*$ for another control function are constructed. And, similarly, for the 
energy we obtain
\be
\label{51}
E_j^* = c_0 \prod_{i=1}^{N_j} \left ( 1 +C_i\; |x|\right )^{m_i} \; .
\ee
The factor approximants (50) and (51) extrapolate the validity of the 
asymptotic series (45) and (47) to the whole region of finite $|x|$. As 
has been shown [47--49], such an extrapolation works very well for the 
whole region of the variable, from zero to infinity, being of especially 
good accuracy for the region between zero and the values of the variable 
of order one. It is exactly the latter region $0\leq|x|\leq|x_c|<1$, 
which we here deal with.

Having in hands the analytical expressions for the sought functions $u_j^*$, 
$v_j^*$, and $E_j^*$, we are already close to our aim of finding an analytical
form of the critical line $x_c(\lbd)$ and, respectively, $g_c(\lbd)$. For this
purpose, it is sufficient to investigate the behaviour of the functions (50) 
and (51), with varying $|x|$ from zero to the point $|x_c|$, where real 
positive solutions stop existing. The critical line $x_c(\lbd)$ defines, 
according to relation (37), the critical line for the coupling
\be
\label{52}
g_c(\lbd) = \frac{(2\pi)^{3/2}}{2\sqrt{\lbd}}\; x_c(\lbd) \; .
\ee
The latter, by definition (18), gives the critical number of atoms $N_c$ 
as a function of the anisotropy parameter $\lbd$.

However, as relations (19) and (20) show, the critical number of atoms 
is not expressed solely through the anisotropy parameter $\lbd$, but also 
involves one of the characteristic lengths (7) or (8). So, $N_c$ is a
function of two parameters, assuming that the scattering length is fixed.
For instance, if we choose the transverse length $l_\perp$ as one of the
parameters and define
\be
\label{53}
\al_\perp(\lbd) \equiv \frac{1}{4\pi}\; |g_c(\lbd)| \; ,
\ee
then the critical number of atoms is
\be
\label{54}
N_c = \al_\perp(\lbd)\left | \frac{l_\perp}{a_s}\right | \; .
\ee
But if the axial length $l_z$ is chosen, then, with the notation
\be
\label{55}
a_z(\lbd) \equiv \frac{\sqrt{\lbd}}{4\pi}\; |g_c(\lbd)| \; ,
\ee
we have the critical number
\be
\label{56}
N_c =\al_z(\lbd) \left | \frac{l_z}{a_s}\right | \; .
\ee
Finally, fixing the length $l_0$, and using the notation
\be
\label{57}
\al_0(\lbd) \equiv \frac{\lbd^{1/6}}{4\pi}\; |g_c(\lbd)| \; ,
\ee
we get the critical number
\be
\label{58}
N_c = \al_0(\lbd) \left | \frac{l_0}{a_s}\right | \; .
\ee
These formulas demonstrate that $N_c$ depends on two 
parameters. Therefore, maximazing one of the critical couplings (53), (55), 
or (57) for finding a maximal $N_c$ implies a conditional variation under 
one of the characteristic  lengths being fixed.

To find the conditional maxima of $N_c$, we need to define the critical 
line $x_c(\lbd)$, where positive solutions for functions (50) or (51) stop 
existing. To estimate the accuracy of the obtained critical values $x_c$, 
we constructed the factor approximants (50) and (51) up to the 17-th order 
and compared the values of $x_c$ for $\lbd=1$ with the exact value (49).
The comparison shows good numerical convergence of the factor approximants,
the sequence $\{ u_j^*\}$ converging a little faster than $\{ E_j^*\}$, 
because of which the values of $x_c$ obtained from the former sequence are 
more accurate. We shall take this into account in what follows, obtaining 
$x_c$ from the most accurate sequence. The error of $x_c$, found from the 
factor approximants up to the 5-th order, is of the order of $10\%$. After 
the fifth-order approximant, the error quickly diminishes to about $1\%$. 
We compared in detail the behaviour of $x_c(\lbd)$ and, respectively, of 
$g_c(\lbd)$, found from the factor approximants up to the fifth order, in 
the whole range of the anisotropy parameter $\lbd$. This behaviour of 
$g_c(\lbd)$ translates into the properties of the related critical couplings 
(53), (55), and (57). Thus, for the transverse coupling (53), the limiting 
behaviour at very small or very large anisotropy parameters is
$$
\al_\perp(\lbd) \simeq 1.253\; \lbd^{1/2} \qquad (\lbd\ra 0)\; ,
$$
\be
\label{59}
\al_\perp(\lbd) \simeq 0.627\; \lbd^{-1/2} \qquad (\lbd\ra\infty)\; .
\ee
For the axial coupling (55), we find the limits
$$
\al_z(\lbd) \simeq 1.253\; \lbd \qquad (\lbd\ra 0)\; ,
$$
\be
\label{60}
\al_z(\lbd) \simeq 0.627 \qquad (\lbd\ra\infty)\; .
\ee
And for the critical coupling (57), we get
$$
\al_0(\lbd) \simeq 1.253\; \lbd^{2/3} \qquad (\lbd\ra 0)\; ,
$$
\be
\label{61}
\al_0(\lbd) \simeq 0.627\; \lbd^{-1/3} \qquad (\lbd\ra\infty)\; .
\ee

Choosing between the critical lines $x_c(\lbd)$, obtained from the factor 
approximants of different orders, we keep in mind our main aim of deriving 
sufficiently simple analytical expressions that would be convenient to study.
The accuracy of the second- and third-order factor approximants are very 
close to each other, with the errors of the order of $10\%$. The accuracy 
of the fourth-order approximant is slightly better, but the expression for
$x_c(\lbd)$ is much more cumbersome. Therefore, we limit ourselves by the 
second-order approximant, which gives the critical line
\be
\label{62}
x_c(\lbd) = -\; \frac{2\lbd}{1+2\lbd} \; .
\ee
Respectively, the critical line for the effective coupling (52) is
\be
\label{63}
g_c(\lbd) = -(2\pi)^{3/2}\; \frac{\sqrt{\lbd}}{1+2\lbd} \; .
\ee
The higher-order versions of the critical lines, with the comparison of 
their properties are given in the Appendix C. From Eq. (53), we obtain the 
radial critical coupling,
\be
\label{64}
\al_\perp =\sqrt{\frac{\pi}{2}}\; \left ( \frac{\sqrt{\lbd}}{1+2\lbd}
\right ) \; ,
\ee
the axial critical  coupling (55),
\be
\label{65}
\al_z =\sqrt{\frac{\pi}{2}}\; \left ( \frac{\lbd}{1+2\lbd}
\right ) \; ,
\ee
and the average coupling (57),
\be
\label{66}
\al_0 =\sqrt{\frac{\pi}{2}}\; \left ( \frac{\lbd^{2/3}}{1+2\lbd}
\right ) \; .
\ee

These simple formulas allow us to accomplish a straightforward search 
for
an optimal trap shape. As is explained above, the result depends on the 
constraint under which the maximal critical number $N_c$ is defined. Thus, 
fixing the radial trap frequency $\om_\perp$, hence, the radial length 
$l_\perp$, and varying the anisotropy parameter $\lbd$, one has to maximize 
the radial critical coupling (64), which gives
$$
\max_\lbd \; \al_\perp(\lbd) = 0.443 \qquad (\lbd=0.5) \; .
$$
This tells us that, under the given conditions, the cigar-shaped trap, 
with $\lbd=0.5$, will be able to confine the maximal number of atoms (54).

If one chooses to fix the axial trap frequency, $\om_z$, hence the axial 
length $l_z$, then one should look for the maximum of the axial critical 
coupling (65), which yields
$$
\max_\lbd \; \al_z(\lbd) = 0.627 \qquad (\lbd\ra\infty) \; .
$$
This means that, under the chosen conditions, the disk-shaped trap can 
contain the maximal number of atoms (56).

And, if one prefers to fix the average trap frequency $\om_0$, hence the 
length $l_0$, then the average critical coupling (66) is to be maximized, 
which leads to
$$
\max_\lbd \; \al_0(\lbd) = 0.418 \qquad (\lbd=1) \; .
$$
Under these conditions, the spherical trap is preferable, if one wishes 
to confine the maximal number of atoms (58).

These conclusions are in agreement with the general physical picture 
discussed in the Introduction. Let us emphasize that if the optimal trap 
shape is considered under a fixed condensate density at the trap center
$$
\rho(0) \sim N \left ( \frac{m\om_0}{\pi}\right )^{3/2} \; ,
$$
that is, under a fixed $\om_0$, then the spherically symmetric trap is 
optimal, in agreement with the above analysis. Note also that fixing the 
density $\rho(0)$, or frequency $\om_0$, is equivalent to fixing the 
condensation temperature
$$
T_c \sim \om_0 \left [ \frac{N}{\zeta(3)}\right ]^{1/3} \; ,
$$ 
where $\zeta(3)$ is the Rieman zeta function.

For the purpose of estimating the critical number of atoms with attractive 
interactions for a given trap with the fixed trap frequencies, that is, with 
the given characteristic lengths, it is convenient to express the critical 
numbers (54), (56), and (58) directly through the trap parameters. Then, 
from any of these expressions, we obtain
\be
\label{67}
N_c =\sqrt{\frac{\pi}{2}}\; 
\frac{l_x\; l_y\; l_z}{|a_s|\;(l_x^2+l_y^2+l_z^2)} \; .
\ee
It is convenient to define the dimensionless quantity
\be
\label{68}
N_c\; \frac{|a_s|}{l_0} = \sqrt{\frac{\pi}{2}} \;
\frac{l_0^2}{l_x^2+l_y^2+l_z^2} \; ,
\ee
where $l_0\equiv(l_xl_yl_z)^{1/3}$. Equation (68) can be rewritten in 
terms of the trap frequencies. Introducing the notation for the average
inverse frequency $\om_{inv}$ through the equality
\be
\label{69}
\frac{1}{\om_{inv}} \equiv \frac{1}{3}\left ( \frac{1}{\om_x} +
\frac{1}{\om_y} + \frac{1}{\om_z} \right ) \; ,
\ee
we find
\be
\label{70}
N_c\; \frac{|a_s|}{l_0} = \sqrt{\frac{\pi}{2}} \; \left (
\frac{\om_{inv}}{3\om_0}\right ) \; .
\ee
Formulas (67) to (70) can serve as a useful tool for a fast and simple 
estimations of the critical number of atoms, with a negative scattering 
length, which could be loaded into a given trap.

\section{Conclusion}

We have considered a dilute gas of trapped Bose atoms at zero temperature, 
when practically all atoms are in the ground state of Bose-Einstein 
condensate. Atoms are interacting through attractive forces, corresponding 
to negative scattering lengths. The number of possible atoms in such a 
Bose-Einstein condensate is limited by a critical number $N_c$ depending 
on the trap parameters. The main result of the present paper is the 
derivation of simple {\it analytical} expressions for estimating this
critical number $N_c$. The investigation of the derived formulas clearly
shows that the number of atoms $N_c$ cannot be expressed through the sole
parameter of anisotropy. Therefore, in order to answer the question "What 
is the optimal trap shape for keeping the maximal number of atoms with 
attractive interactions", it is necessary to specify under what conditions 
one is looking for this maximum. Thus, if the radial trap frequency 
$\om_\perp$ is kept fixed and the anisotropy parameter $\lbd$ is varied, 
then the cigar-shaped trap with $\lbd=0.5$ is optimal. But if the axial 
trap frequency $\om_z$ is fixed, then the disc-shaped trap is optimal. And 
when the geometric-average frequency $\om_0$ is fixed, with the anisotropy 
parameter being varied, then the spherically symmetric trap becomes optimal. 
The spherical trap is also optimal, when the condensate density is kept 
fixed. In general, for estimating the critical number of atoms $N_c$, it is 
convenient to use formulas (67) to (70), which are represented in the form 
valid for arbitrary trap frequencies $\om_x$, $\om_y$, and $\om_z$, hence, 
for arbitrary trap lengths $l_x$, $l_y$, and $l_z$. It is easy to check that 
these estimates are in good agreement with experiment.

For instance, in the experiments [12,50] with $^7$Li, having the negative
scattering length $a_s=-27.3a_B=-14.5\AA$, where $a_B$ is the Bohr radius, 
the trap with the frequencies $\om_x=2\pi\times 150.6$ Hz, $\om_y=2\pi\times 
152.6$ Hz, and $\om_z=2\pi\times 131.5$ Hz was used. This, for the mass 
$m_{\rm Li}=11.6\times 10^{-24}$ g, translates into the characteristic 
lengths $l_x=0.309\times 10^{-3}$ cm, $l_y=0.307\times 10^{-3}$ cm, and 
$l_z=0.331\times 10^{-3}$ cm. In these experiments [12,50] the critical 
number was estimated being between 600 and 1300 atoms, that is, on average, 
$N_c\approx 950$. Our formula (67), for this case, gives $N_c=910$.

In the experiments [14,15] with $^{85}$Rb, the scattering length could be 
varied by means of the Feshbach resonance techniques. The trap had the 
frequencies $\om_x=2\pi\times 17.24$ Hz, $\om_y=2\pi\times 17.47$ Hz, and 
$\om_z=2\pi\times 6.80$ Hz, that is $\om_0=2\pi\times 12.7$ Hz. The 
dimensionless parameter (70) was found to be equal to $N_c|a_s|/l_0=0.46
\pm 0.07$. From our formula (70) for this trap, we find $N_c|a_s|/l_0=0.47$.

In this way, the derived formulas allow one to quickly get rather accurate 
estimates for the critical number of atoms. The advantage of these formulas 
is their simple analytical representation. Therefore, to find $N_c$ and 
to optimize the trap shape, one does not need to plunge into lengthy and 
complicated numerical calculations, but it is sufficient to employ the 
derived simple analytical formulas.

\vskip 5mm

{\bf Acknowledgement}

\vskip 2mm

We are grateful for the financial support to the German Research 
Foundation.

\newpage

{\large{\bf Appendix A. Properties of Optimized Approximants}}

\vskip 5mm

The control functions and the optimized approximant for the dimensionless 
energy, discussed in Sec. III, have the following asymptotic behaviour 
in the weak-coupling and strong-coupling limits.

In the weak-coupling limit, when $g\ra\pm 0$, the effective coupling 
parameter (27) is such that $x\ra\pm 0$. Then from the optimization 
equations (31), we find
$$
u \simeq 1  -\; \frac{1}{2p}\; x + \frac{p+3q}{8p^2q}\; x^2 \; - \;
\frac{3p^2+16pq+20q^2}{64p^3q^2}\; x^3 \; ,
$$
$$
v \simeq 1  -\; \frac{1}{2q}\; x + \frac{p+q}{4pq^2}\; x^2 \; - \;
\frac{7p^2+20pq+12q^2}{64p^2q^3}\; x^3 \; .
$$
These expressions satisfy the boundary conditions (32).

In the strong-coupling limit, when $g\ra\infty$ and $x\ra\infty$, we 
have the control functions
$$
u \simeq \left ( \frac{p^3}{q}\right )^{1/5} x^{-2/5} +
\frac{q^2-3p^2}{5(pq^3)^{1/5}}\; x^{-6/5} + 
\frac{3p^4-p^2q^2-q^4}{5pq}\; x^{-2}\; ,
$$
$$
v \simeq \left ( \frac{q^2}{p}\right )^{2/5} x^{-2/5} +
\frac{2(p^2-2q^2)}{5}\left (\frac{q}{p^3}\right )^{2/5}\; x^{-6/5} + 
\frac{6q^4-4p^2q^2-p^4}{5p^2}\; x^{-2}\; .
$$
The strong-coupling limit for attractive interactions, for which 
$x\ra-\infty$, is not defined, since the solutions to Eqs. (31) become 
complex.

The weak-coupling limit for the optimized energy (35) can be written as 
an expansion
$$
\tilde E \simeq a_0 + a_1 x + a_2 x^2 + a_3 x^3 \qquad (x\ra \pm 0)
$$
in which the coefficients are
$$
a_0 = p+\frac{q}{2}\; , \qquad a_1 =\frac{1}{2} \; , \qquad 
a_2 = -\; \frac{p+2q}{16pq}\; , \qquad a_3 =\left ( \frac{p+2q}{8pq}
\right )^2 \; .
$$
Respectively, the strong-coupling limit for the energy can be 
represented as
$$
\tilde E \simeq b_0\; x^{2/5} + b_1\; x^{-2/5} + b_2\; x^{-6/5} +
b_3\; x^{-2} \; ,
$$
where the coefficients are
$$
b_0 = \frac{5}{4}\left (p^2q\right )^{1/5} \; , \qquad b_1 =
\frac{2p^2+q^2}{4(p^2q)^{1/5}} \; ,
$$
$$ 
\quad b_2 = -\; \frac{3p^4-2p^2q^2+2q^4}{20(p^2q)^{3/5}} \; , 
\qquad b_3 = \frac{2p^6-p^4q^2-2p^2q^4+2q^6}{20p^2q} \; .
$$
These expansions are valid for arbitrary energy levels. The quantum 
numbers enter through notation (30).

\newpage

{\large{\bf Appendix B. Expansions for Attractive Interactions}}

\vskip 5mm

For atoms with attractive interactions, it is convenient to use the 
relation $x=-|x|$ and to express all quantities as functions of $|x|$. 
In expression (45) for the control function $u$, the coefficients are
$$
u_0 = 1 \; , \qquad u_1 =\frac{1}{2}\; , \qquad u_2 =
\frac{1+3\lbd}{8\lbd}\; , \qquad 
u_3 = \frac{(1+2\lbd)(3+10\lbd)}{64\lbd^2} \; , 
$$
$$ 
u_4= \frac{2+20\lbd+48\lbd^2+35\lbd^3}{128\lbd^3} \; , \qquad
u_5 = \frac{15+336\lbd+1400\lbd^2+2048\lbd^3+1008\lbd^4}{4096\lbd^4} \; ,
$$
$$
u_6 =\frac{(1+\lbd)(35+221\lbd+409\lbd^2+231\lbd^3)}{1024\lbd^4}\; .
$$
Expansion (46) for the control function $v$ has the coefficients
$$
v_0=1 \; , \qquad v_1 =\frac{1}{2\lbd}\; , \qquad v_2 =
\frac{1+\lbd}{4\lbd^2}\; , \qquad
v_3 = \frac{7+20\lbd+12\lbd^2}{64\lbd^3} \; ,
$$
$$
v_4= \frac{5+32\lbd+48\lbd^2+20\lbd^3}{128\lbd^4} \; , \qquad
v_5 = \frac{39+616\lbd+1800\lbd^2+1792\lbd^3+560\lbd^4}{4096\lbd^5} \; ,
$$
$$
v_6 =\frac{35+192\lbd+350\lbd^2+256\lbd^3+63\lbd^4}{512\lbd^5}\; .
$$
And in expansion (47) for the optimized energy, the coefficients are
$$
c_0=\frac{2+\lbd}{2} \; , \qquad c_1 =-\;\frac{1}{2}\; , \qquad 
c_2 = -\; \frac{1+2\lbd}{16\lbd}\; , \qquad
c_3 =-\; \frac{(1+2\lbd)^2}{64\lbd^2} \; ,
$$
$$
c_4=-\; \frac{1+8\lbd+16\lbd^2+10\lbd^3}{256\lbd^3} \; , \qquad
c_5 =-\; \frac{3+56\lbd+200\lbd^2+256\lbd^3+112\lbd^4}{4096\lbd^4} 
\; ,
$$
$$
c_6 =-\;\frac{(1+\lbd)^2(5+22\lbd+21\lbd^2)}{1024\lbd^4}\; .
$$
These expansions are written here for the ground-state level, which looses 
stability before other states with higher energies for $x\ra x_c$ and 
$g\ra g_c$, when attractive interactions become too strong. This is easy 
to understand keeping in mind that with increasing $|x|$ or $|g|$, that is, 
decreasing $x$ and $g$, all energy levels move down. Consequently, the lowest 
energy level will be the first to touch zero and then become complex, 
provoking the system instability.

\newpage

{\large{\bf Appendix C. Properties of Critical Lines}}

\vskip 5mm

The critical line $x_c(\lbd)$ is obtained as is explained in Sec. IV, from 
analysing the behaviour of the related self-similar factor approximants. 
On the line $x_c(\lbd)$, the ground-state energy becomes complex-valued. 
Generally, for a $k$-th order factor approximant, we get a $k$-th order 
critical line $x_c^{(k)}(\lbd)$ and, respectively, the effective-coupling 
critical line
$$
g_c^{(k)}(\lbd) = \frac{(2\pi)^{3/2}}{2\sqrt{\lbd}}\; x_c^{(k)}(\lbd) \; .
$$
In the second order, we have
$$
x_c^{(2)}(\lbd) = -\; \frac{2\lbd}{1+2\lbd} \; ,
$$
which is the critical line (62). In the third order, we get
$$
x_c^{(3)}(\lbd) = -\; \frac{16\lbd}{9+16\lbd} \; ,
$$
which is very close to $x_c^{(2)}$. And for the fourth order, we find
$$
x_c^{(4)}(\lbd) = -\;
\frac{8\lbd(1+4\lbd)}{\sqrt{2(19+36\lbd+56\lbd^2+64\lbd^3)}-
2+20\lbd+32\lbd^2} \; .
$$
The higher-order expressions $x_c^{(k)}(\lbd)$ become quickly so much 
cumbersome that the whole idea of deriving simple analytical expressions, 
allowing for an easy analysis, looses grounds. This is why, in the main 
text of the paper the simplest forms for the critical lines (62) and (63), 
corresponding to the second-order approximations, were used. In addition, 
the functions $g_c^{(k)}(\lbd)$ of all orders demonstrate rather 
similar behaviour. For instance, for the absolute values of the critical 
lines $g_c^{(k)}(\lbd)$, we obtain the following asymptotic behaviour in 
the limits of the small and large anisotropy parameter $\lbd$.

In the limit $\lbd\ra 0$, related to a very elongated cigar-shaped trap, 
we have
$$
|g_c^{(2)}(\lbd)| \simeq 15.7496\; \lbd^{1/2} - 31.4992\; \lbd^{3/2} + 
62.9984\; \lbd^{5/2} \; ,
$$
$$
|g_c^{(3)}(\lbd)| \simeq 13.9997\; \lbd^{1/2} - 24.8883\; \lbd^{3/2} +    
44.2458\; \lbd^{5/2} \; , 
$$
$$
|g_c^{(4)}(\lbd)| \simeq 15.1278\; \lbd^{1/2} - 33.3560\; \lbd^{3/2} +  
67.7766\; \lbd^{5/2} \; .
$$

In the opposite limit $\lbd\ra\infty$, which refers to a very plane 
disk-shaped trap, we obtain
$$
|g_c^{(2)}(\lbd)| \simeq 7.8748\; \lbd^{-1/2} - 3.9374\; \lbd^{-3/2} +  
1.9687\; \lbd^{-5/2} \; ,
$$
$$
|g_c^{(3)}(\lbd)| \simeq 7.8748\; \lbd^{-1/2} - 4.4296\; \lbd^{-3/2} +
2.4916\; \lbd^{-5/2} \; ,
$$
$$
|g_c^{(4)}(\lbd)| \simeq 7.8748\; \lbd^{-1/2} - 1.9687\; \lbd^{-3/2} +
1.7226\; \lbd^{-5/2} \; .
$$
Note that the limits
$$
\lim_{\lbd\ra\infty}\; x_c^{(k)}(\lbd) = -1 \; ,
$$
$$
\lim_{\lbd\ra\infty}\; \sqrt{\lbd}\; |g_c^{(k)}(\lbd)| = 
\frac{(2\pi)^{3/2}}{2} 
$$
are the same in all orders.

Looking for the optimal trap shape at a fixed condensate density, that is, 
under a fixed frequency $\om_0$, we have to study the behaviour of the 
effective coupling
$$
\al_0^{(k)}(\lbd) = \frac{\lbd^{1/6}}{4\pi} \; \left | 
g_c^{(k)}(\lbd)\right | = \frac{\sqrt{2\pi}}{4\lbd^{1/3}}\;
\left | x_c^{(k)}(\lbd)\right | \; .
$$
The maximum of this quantity defines the critical number (58). For all 
orders $k$, the maximum of $\al_0^{(k)}(\lbd)$ happens at $\lbd\approx 1$. 
Thus, $\lbd^{(2)}=1$, $\lbd^{(3)}=1.1$, and $\lbd^{(4)}=0.9$. This means
that, under a fixed condensate density, the spherical trap is optimal.

\end{document}